# Magnetic Properties of $NH_4H_2PO_4$ and $KH_2PO_4$: Emergence of Multiferroic Salts

Lei Meng,[1#] Chen He,[1#] Wei Ji,[2] Fei Yen[1*]

[1]State Key Laboratory on Tunable Laser Technology, Ministry of Industry and Information Technology Key Laboratory of Micro-Nano Optoelectronic Information System and the School of Science, Harbin Institute of Technology, Shenzhen, University Town, Shenzhen, Guangdong 518055, P. R. China
[2]Education Center of Experiments and Innovations, Harbin Institute of Technology, Shenzhen, University Town, Shenzhen, Guangdong 518055, P. R. China

**Abstract:** We observe sharp step-down discontinuities in the magnetic susceptibility of $NH_4H_2PO_4$ and $NH_4H_2PO_4$-$d_{60}$ (60% deuterated) along the *a* and *c*-axes occurring exactly at their antiferroelectric transition temperatures. For the case of $KH_2PO_4$, less pronounced discontinuities occur at the ferroelectric transition temperature. To explain this, we treat the acid protons as individual oscillators that generate current elements which translate to magnetic forces in near resonance with each other. With decreasing temperature, the resonant forces become more commensurate which amplifies a disproportionate drop off of two types of magnetic forces to eventually trigger the structural phase transitions. For the case of $NH_4H_2PO_4$, the associated internal magnetic field appears to aid the $NH_4^+$ to order at higher temperature. At 49 K, a shoulder-like anomaly in both $NH_4H_2PO_4$ and $KH_2PO_4$ is attributed to a possible onset of macroscopic quantum tunneling of protons. Our findings bring forth a new category of intrinsic multiferroic systems.

**ToC Graphic:**

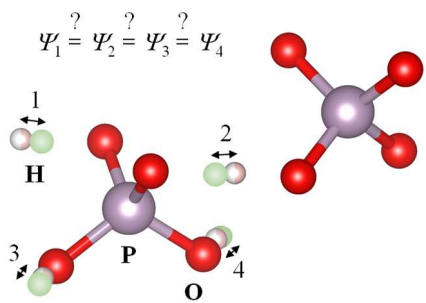

Most medicines, energy conversion and energy storage systems possess hydrogen atoms which are believed to play the central role in their functionalities.[1-3] Yet, as to how these protons (H$^+$) formulate to manifest the observed macroscopic properties remains ever elusive. All of the electrons in these hydrogen-based compounds reside within closed orbital shells so a large portion of the input thermal energy is channeled onto the next lightest particles rendering the protons highly energetic. Consequently, most hydrogen-bonded systems are diamagnetic which in theory should have a magnetic susceptibility of $\chi_e = \mu_0 \cdot \mu_e \cdot N / H$ (where $\mu_0$, $\mu_e$, $N$ and $H$ are the permeability of free space, magnetic moment of the electron, number of atoms per unit volume and applied magnetic field, respectively).[4] Not only does this constant not hold true for most hydrogen-based materials, the susceptibility for many also vary with temperature.[5-7] This can mean that either the motion of protons disrupts the diamagnetic response of the electrons or the protons have an orbital magnetic moment of their own. In either case, most models treat the protons as rigid and continue to overanalyze the behavior of the electrons which are all paired up so their effects cancel each other out. Although it is fair to dismiss the protons as static charges since their motions always bring them back to their original equilibrium positions leaving the lattice essentially unchanged, during the process of traveling from one equilibrium point to another, a magnetic field is generated. Recently, we showed how the orbital motion of protons in reorienting NH$_4^+$ tetrahedra at rates of $10^{12}$ Hz generates a magnetic moment that is larger than the intrinsic spin of the proton.[6] The magnitude of these orbital moments is nearly three orders of magnitude smaller than that of an electron, however, in the proton disordered state, the proton moments can only point along a discrete number of directions due to the four-fold nature of the potentials confining each NH$_4^+$. This is in great contrast to the spin (and orbit) orientation of electrons in the paramagnetic state where they can point along a continuous array of directions. As such, proton orbital-orbital interactions are actually greatly enhanced to the point that we concluded that the low temperature structural phase transitions in the ammonium halides are a consequence of magnetic ordering of NH$_4^+$.[6] The way the protons are assembled in hydrogen-based compounds come in many forms other than tetrahedrally shaped moieties such as CH$_4$ or BH$_4^-$. The reorienting functioning group may possess an intrinsic dipole moment such as $CH_3NH_3^+$ or the protons may simply go it alone and hop around throughout a multitude of available empty sites.[8] In this work, we study potassium dihydrogen phosphate (KDP) and ammonium dihydrogen phosphate (ADP) because the acid protons only possess two equilibrium points. It will be shown that even in this simplest case, through measurements of the magnetic susceptibility, we find evidence that magnetic interactions render the system ferroelectric. By implementing the notion that periodic motion of protons generates an alternating internal magnetic field that allows them to interact with each other depending on the degree of coherence, researchers will obtain a better understanding of the fundamental causes underlying the many functionalities of hydrogen-based compounds. For example, there is still no general consensus on explanations of why the methylammonium lead halide perovskites possess such high photovoltaic conversion rates[1] or why some borohydrides possess enhanced ionic conductivities in their proton disordered phases.[2]

Potassium dihydrogen phosphate (KDP) $KH_2PO_4$ and ammonium dihydrogen phosphate (ADP) $NH_4H_2PO_4$ are renowned for their nonlinear optical properties and are widely employed in nonlinear optical and electro-optic devices.[9] At room temperature, both are tetragonal (Space group $I\bar{4}2d$)[10,11] and at $T_C$ = 122.7 K KDP becomes ferroelectric[12] while ADP becomes antiferroelectric at $T_N$ = 148.2 K.[13] Both low temperature phases are orthorhombic with the main difference being that the *a*- and *b*-axes of KDP rotate by 45° while remaining unchanged in ADP.[14] Despite having its own category in the subfield of ferroelectricity, the fundamental cause underlying the spontaneous polarization of the KDP family remains unclear: all that can be said is that ferroelectricity is due to the ordering of hydrogen bonds, its microscopic origin remains unknown. One feature researchers agree upon is that it is the acid protons, the $H^+$ of the $H_2PO_4^-$ groups, that somehow "trigger" the structural transition to the orthorhombic phase.[15] Some studies have revealed that these acid protons actively jump back and forth between two symmetrical sites along the O–H···O paths.[16-19] Recently, the ferroelectric phase of ammonium sulfate $(NH_4)_2SO_4$ was identified to be the result of magnetic ordering of two inequivalent types of proton orbitals[20] which simultaneously break the time-reversal and spatial-inversion symmetries, the one requirement of a multiferroic system. In the current system, the $PO_4^{3-}$ tetrahedra are slightly elongated along the *a-b*-axes so the back and forth oscillating of the acid protons is also divided into two groups which would also break both time-reversal and spatial-inversion symmetries should they become magnetically ordered. Unlike reorienting $NH_4^+$ which carry an orbital magnetic moment, the acid protons do not follow circular paths so a different microscopic model is necessary. Figure 1a shows the two sites an acid proton can reside in for KDP at 5° above its $T_C$ according to neutron diffraction measurements.[11] The time it takes to travel from one site to the other is ~$10^{-12}$ s.[17] With a distance of Δx=0.343 Å, the maximum speed is ~34.3 m/s. We make the following two assumptions: 1) the protons are not stationary, so they continuously jump back and forth between two sites[16] and 2) the process is mostly classical in nature above 50 K because the jump rates are proportional with temperature,[21,22] meaning that the hops are mainly caused by thermal agitation rather than zero-point energy. Since a moving charge equates to a current, the periodic motion of the proton generates a sinusoidal alternating current with an amplitude of $I = dq/dt = +q \cdot f \approx 1.602 \times 10^{-7}$ A, (where $q = 1.602 \times 10^{-19}$ C is the elemental unit of charge, $f = 10^{12}$ Hz). Figure 1b shows a magnetic field line associated to the current produced by the proton at position 3. At a distance of $r_{23}$ = 3.517 Å away, the average distance of the other proton of the $H_2PO_4^-$ ion situated at position 2, the amplitude of the magnetic field is $\mathbf{B} = \mu_0 \mathbf{I}/2\pi \cdot r = 0.91 \times 10^{-4}$ T ≈ 1 Gauss (where $\mu_0 = 12.57 \times 10^{-7}$ H/m). The associated force is $\mathbf{F} = q \cdot \mathbf{v} \cdot \mathbf{B} \sin\theta = 5.0 \times 10^{-22}$ N with θ as the angle a field line makes with the direction of oscillation. This figure is rightfully small but its effects are amplified because the frequency of the driving force of the proton at position 3 is near the same as the resonant frequency of the proton at position 2 and vice versa. In fact, the electronegativity of $H_2PO_4^-$ dictates that the resonant frequencies of its acid protons be identical in the ground state.[23] At elevated temperatures, the coherence of the two protons may be slightly off, however, we posit that with decreasing temperature the frequencies of the two acid protons become more and more commensurate to eventually render the system unstable and trigger the structural transition to the orthorhombic phase. From such, we employ the universal driven harmonic oscillator equation to describe this amplification effect:

$$\Delta d = \sum_{i,j=1,\ i\neq j}^{4} \frac{F_{ij}(T) \sin \phi_{ij}}{m_p \sqrt{\left(2\omega_i(T)\omega_{0j}(T)\xi\right)^2 + \left(\omega_{0j}(T)^2 - \omega_i(T)^2\right)^2}}$$

where $\Delta d$ are small additional displacements perpendicular to the direction of oscillations along the $ab$-plane, $m_p$ is the mass of the proton, $\omega_i(T)$ is the frequency of the driving force (which is proportional to $f^2$ so it is temperature dependent) $F = F_{ij}(T)$ of the proton at position $i$, $\omega_{0j}(T)$ the natural frequency of the proton at position $j$ and $\xi$ a damping ratio which we consider to be negligible. The average distance between protons at positions 1 and 2 $r_{12}$ is larger than $r_{13}$ and $r_{14}$ due to the distorted nature of the $PO_4^{3-}$ so $F_{12}(T)$ is much smaller than $F_{13}(T)$ and $F_{14}(T)$. At the same time, the angle the magnetic field lines make with the direction of the current between each of the four protons are also different with $\phi_{12}$ much smaller than $\phi_{13}$ and $\phi_{14}$. While $F_{ij}(T)$ clearly decreases with cooling, $\omega_i(T)$ and $\omega_{0j}(T)$ also become more coherent which renders Equation (1) to diverge and eventually trigger the observed structural phase transition. Since the internal magnetic field changes during the process, a small change in the magnetization should be detected in KDP at $T_C$. For the case of ADP, the internal magnetic field generated by the acid protons should aid the magnetic moments of the reorienting $NH_4^+$ to order so pronounced discontinuities are expected at $T_N$. To verify these hypotheses, we measured the magnetic susceptibility of ADP and KDP presented below.

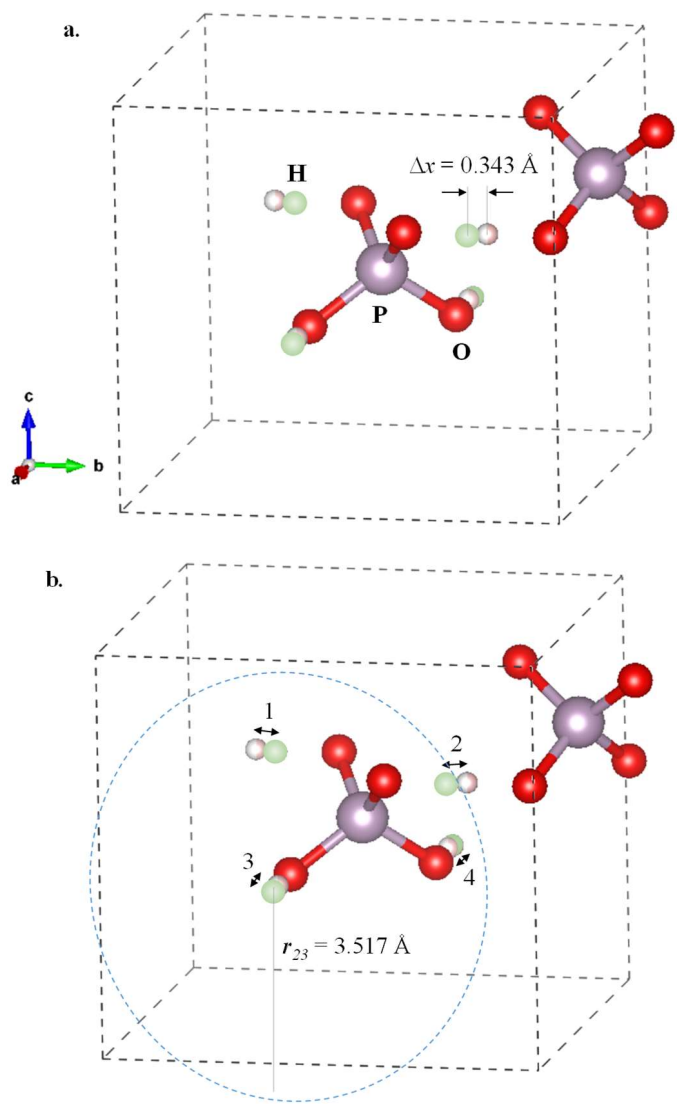

**Figure 1: a.** Two PO$_4^{3-}$ tetrahedra linked by a hydrogen atom via an O–H···O path in the paraelectric phase at 127 K after Nelmes *et al.*[11] Each proton (H$^+$) can reside in one of two available sites separated by $\Delta x$ with equal probability. **b.** The protons jump back and forth between their two available sites which generate a current element. The shortest average distance between two protons is $r_{23}$ = 3.517 Å. The blue dashed line is a magnetic field line generated by the proton at position 3 which mainly lies in the *b-c*-plane. The current element of the proton at position 2 points mainly along the ±*b*-axis, hence, an additional force element along the ±*a*-axis exists that is proportional to the jump rates of the protons. This results in small additional displacements $\Delta d$ along directions perpendicular to the oscillations that are highly dependent of the degree of coherence of the proton motions according to Equation (1).

ADP and KDP both 99.99% in purity were acquired from Aladdin-e, Inc. (Shanghai) and mixed with deionized H$_2$O to make slightly oversaturated solutions. After slow evaporation of the solution, clear and transparent crystals of dimensions around 1.5 x 2 x 5 mm emerged. The axes were identified from their growth habits and double checked with measurements of the dielectric constant. For the deuterated samples, heavy water was employed as the solution. The magnetic susceptibility was measured by the vibrating sample magnetometer option of a PPMS (Physical Properties Measurement System) Dynacool manufactured by Quantum Design, Inc. (San Diego).

Figure 2 shows the molar magnetic susceptibility $\chi_c(T)$ with respect to temperature of ADP along the *c*-axis orientation under an applied magnetic field of $H = 10$ kOe. From the slope of the magnetization *M* vs. *H* curve (inset of Fig. 2), $\chi_c(T) = -87 \times 10^{-6}$ mol/cm$^3$ at 296 K. This is noted because we were not able to find an experimental figure for ADP and KDP in the existing literature. This is probably because these values have always been presumed to be uninteresting as they are expected in theory to remain constant all the way down to absolute zero as shown by the horizontal dotted line $\chi_e$. The feature we are most interested on are the sharp step anomalies observed between the paraelectric and antiferroelectric phase transitions. Figure 3 is an enlarged view near the transition temperature for the case when $H = 50$ kOe was applied along the *c* and *a*-axis. The critical temperatures are $T_{N\_c} = 147.0$ K during cooling and $T_{N\_w} = 149.3$ K during warming, in excellent agreement with heat capacity[24] and dielectric constant experiments.[25] The hysteresis of 2.1 K is also in par with that observed in most other experiments summarized in Ref. [25]. A second newly identified feature is a shoulder-type of anomaly at $T_p = 49$ K which becomes more pronounced during warming. At temperatures above ~15 K the cooling and warming curves do not retrace each other which is also consistent with reports on measurements performed on other physical parameters.[25] Below 15 K, both curves are almost identical and proportional to $T^{-0.5}$ after subtracting away the background diamagnetic contribution by the electrons $\chi_e$.

To make sure the observed features are inherent of the protons, we performed a deuteration process on ADP which replaced 60% of the hydrogen atoms by deuterons. Figure 4 shows $\chi(T)$ of NH$_4$H$_2$PO$_4$-$d_{60\%}$ (at 60% deuteration) from 5 – 300 K under *H*

= 10 kOe along the *c*-axis. The inset is an enlarged view near the transition point and shows the behavior of χ(*T*) when cycled back and forth between 180 and 225 K. The same step anomalies were observed as in ADP, however, they shifted to $T_{N60\_c}$ = 201.0 K during cooling and $T_{N60\_w}$ = 203.3 K when warming. The values of the critical temperatures are in good agreement with the well-known isotope effect of ADP[13,14] where the structural phase transition increases linearly from 149 K to 242 K when the sample is fully hydrogenous to 98% deuterated, respectively. The electronic configurations of the two samples may be treated as being nearly identical to each other so the shifting of the critical temperatures in the magnetic susceptibility is more evidence that the structural phase transitions are due to the protons with the added feature that the observed magnetic behavior stems from protons (and deuterons).

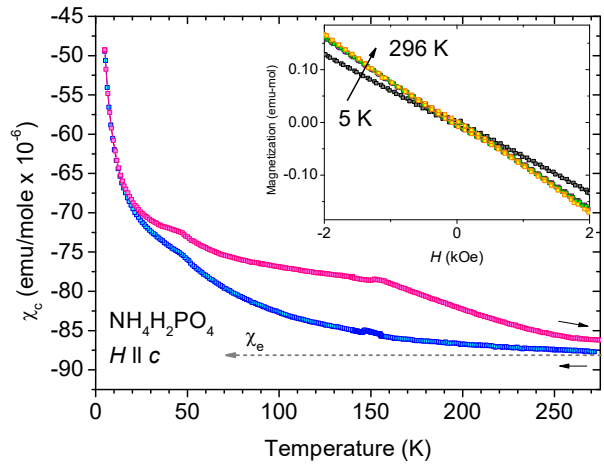

**Figure 2:** Magnetic susceptibility vs. temperature χ(*T*) of ammonium dihydrogen phosphate NH$_4$H$_2$PO$_4$ (ADP) under magnetic field of *H* = 10 kOe oriented along the *c*-axis direction. The directions of the arrows indicate the warming (red) and cooling (blue) curves. Inset shows the magnetization vs. *H* at 5, 125, 170 and 296 K. $χ_e$ is the contribution by the electrons shown as the horizontal dotted line.

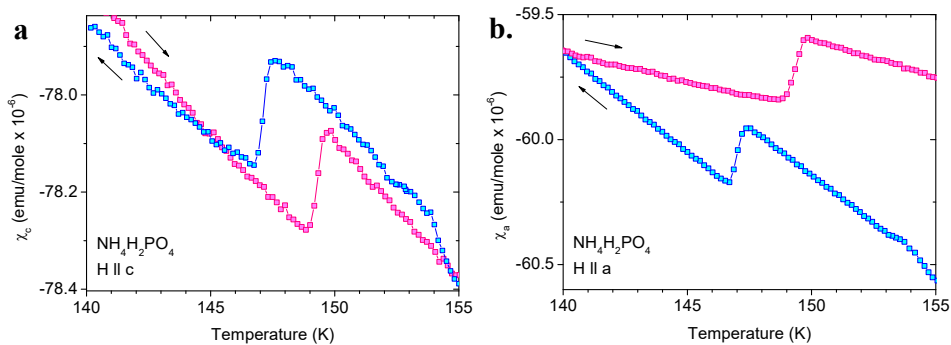

**Figure 3:** Enlarged views of χ(*T*) along the **a.** *c*- and **b.** *a*- axes at *H* = 50 kOe. The step anomalies occur at $T_{N\_c}$ = 147.0 K and $T_{N\_w}$ = 149.3 K during cooling and warming, respectively.

Figure 5 shows $\chi_c(T)$ for the case of KDP from 5 – 300 K also under *H* = 10 kOe. The molar susceptibility at room temperature is 80 x $10^{-6}$ mol/cm³. Instead of the sharp step anomalies as in ADP, minute peaks are apparent during cooling at $T_{C\_c}$ = 122.0 K and $T_{C\_w}$ = 123.3 K during warming only when *H* is applied along the *b*-axis (inset of Fig. 5). These critical temperatures are in good agreement with heat capacity[26] and dielectric constant measurements.[12] The absence of sharp step-anomalies along the *c*-axis direction further substantiates that the reorienting $NH_4^+$ in ADP indeed possess a magnetic moment. The shoulder-like anomaly at $T_p$ = 49 K in ADP is also apparent in KDP. Also, below 16 K both the cooling and warming curves begin to diverge and retrace each other just like in the case of ADP.

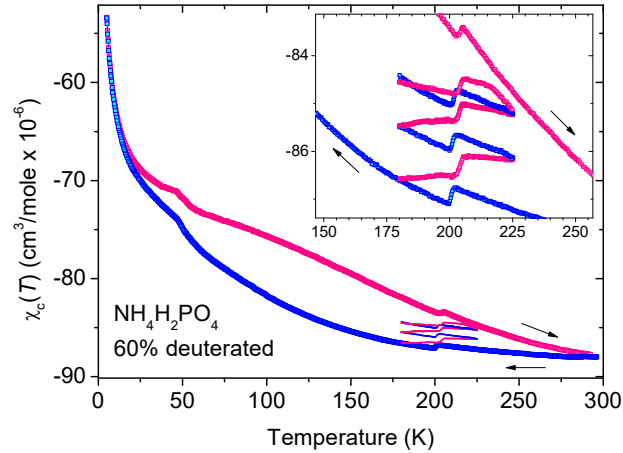

**Figure 4:** χ(*T*) of ammonium dihydrogen phosphate deuterated at 60% at *H* = 10 kOe oriented along the *c*-axis direction. The transition temperatures coincide with the large isotope effect where the fully deuterated analogue $ND_4D_2PO_4$ has a critical temperature of 242 K.

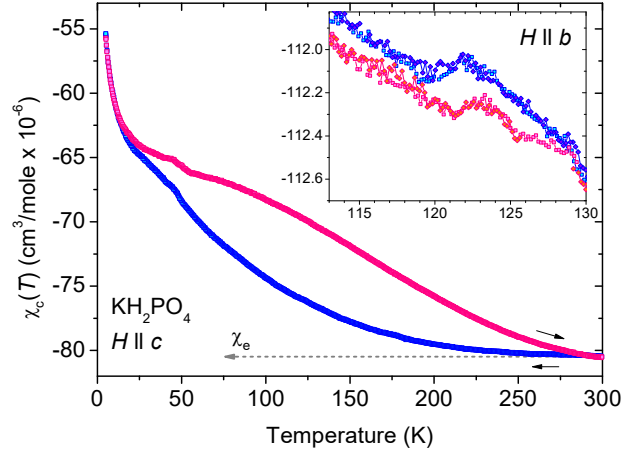

**Figure 5:** $\chi(T)$ of $KH_2PO_4$ (KDP) under $H = 10$ kOe along the $c$-axis direction. Inset: enlarged view of two subsequent runs near $T_C$ under $H = 90$ kOe applied along the $b$-axis direction.

In solid-solid phase transitions, discontinuities appear in $\chi(T)$ *only* when there is magnetic ordering; no anomalies appear even when a material melts. From such, the sharp step anomalies in ADP is concrete evidence that the antiferroelectric transition is magnetically-driven by the ordering of $NH_4^+$. On the other hand, the anomalies at $T_C$ in KDP are more subtle because there are no $NH_4^+$ to order. Nevertheless, there still exist minute anomalies upon crossing $T_C$ and starting from room temperature down to 5 K, $\chi(T)$ decreases by nearly 40% with the existence of an additional anomaly at 49 K. Since the diamagnetic contribution from the electrons is constant, the changes in $\chi(T)$ reflect the changes in $B$ generated by the current elements of the acid protons. From such, we posit that the structural phase transition stems from a disproportionate amplification of forces according to Equation (1) due to the resonant frequencies becoming more and more coherent with cooling. The force experienced by the acid protons is along a direction perpendicular to their oscillations which explains the shearing of the $a$-axis and $b$-axis by 45° at $T_C$.[14]

If we treat the orbital angular momentum $L_p$ of a proton revolving about a nitrogen atom as being quantized, the lowest energy it can have is 30 cm$^{-1}$ when $L_p = m_p \cdot r_p \times v = 1\,\hbar$ (where $r_p = 0.731$ Å is the radius of the proton orbital). The temperature of the equivalent thermal energy is 47 K so it appears that $T_p$ is a transition temperature between the classical and quantum regimes where the protons collapse to their ground tunneling states. Whether a proton crosses a potential barrier by hopping over it or tunneling through it, the physically path is the same on both instances. The main difference lies in the energy distribution. In the classical scenario, protons mainly employ thermal energy to overcome the potential barrier which have values near $k_BT$. In the quantum scenario, protons can be perceived as first borrowing energy from the lattice which is then paid back once the tunneling process is over. The latter process requires a high degree of correlation between the protons, something not possible when $T > T_p$ due to thermal disruptions. A possible mechanism is the pairing of protons which have spin 1/2 to add up to become a composite boson of spin 1 very much like Cooper pairs. This allows the protons to access the same ground tunneling state and become highly correlated. We believe this is a necessity in order for the system to deal with the presence of zero-point energy and find it appropriate since helium atoms, four times more massive than protons, overcome this dilemma by forming a superfluid. Hence, with thermal noise becoming less and less at low temperature and $\bm{F}_{ij}(T)$, $\omega_{0j}(T)$, and $\omega_i(T)$ becoming independent of temperature in the ground state, the observed $T^{-0.5}$ dependence in $\chi(T)$ appears to be the consequence of macroscopic quantum tunneling which is believed to be capable to "enormously amplify" the effects of magnetic interactions according to Leggett.[27]

In conclusion, we observe pronounced anomalies in the magnetic susceptibility at the Néel temperature of ADP. For the case of KDP, the discontinuities at its Curie temperature are less pronounced. In both cases, the susceptibility decreases by nearly 40% from room temperature down to 5 K and a shoulder-like anomaly is observed at 49 K. These results provide evidence that resonant magnetic interactions dominate the molecular dynamics of the system: from the triggering of the triclinic to orthorhombic structural phase transitions to the possible emergence of a previously unidentified type of macroscopic quantum phenomena. The protons in essentially all hydride compounds exhibit periodic motion of some sort. A subset of these systems possesses two or more inequivalent types of orbital motion which upon their magnetic order leads to lattice distortions containing off-center sites. From such, our findings open up an entire new category of multiferroic systems.

**Author Information:**

#L. Meng and C. He contributed equally.

Corresponding Author:

*(F. Y.) E-mail: fyen@hit.edu.cn. Tel: +86-1334-929-0010.

ORCID:

Fei Yen: 0000-0003-2295-3040

Notes:

The authors declare no competing financial interests.

**Acknowledgements:**

F. Y. is grateful of a General Project grant of the Shenzhen Universities Sustained Support Program and a Project Initiative Fund No. 20188014 of the School of Science of the Harbin Institute of Technology, Shenzhen.